\documentclass[letterpaper, 10 pt, journal, twoside]{IEEEtran}
\usepackage[pdftex]{graphicx} 
\usepackage{gensymb}
\usepackage{amsmath}
\usepackage{hyperref}
\IEEEoverridecommandlockouts  
\usepackage{booktabs}
\usepackage{xcolor}
\usepackage[normalem]{ulem}
\usepackage{tikz}
\usepackage{xfrac}
\usepackage{float}
\usepackage{placeins}
\usepackage[justification=centering]{caption}
 \usepackage[pscoord]{eso-pic}
 \usepackage{fancyhdr}
\usepackage{atbegshi}

\begin{document}
\title{Artificial Intelligence-based Analysis of Change in Public Finance between US and International Markets}
\author{
\begin{tabular}{c} Kapil Panda \\ University of North Texas \\ Denton, TX, United States of America \\ kapilpanda@my.unt.edu \end{tabular}
}
\maketitle

\begin{abstract}
Public finances are one of the fundamental mechanisms of economic governance that refer to the financial activities and decisions made by government entities to fund public services, projects, and operations through assets. In today's globalized landscape, even subtle shifts in one nation's public debt landscape can have significant impacts on that of international finances, necessitating a nuanced understanding of the correlations between international and national markets to help investors make informed investment decisions. Therefore, by leveraging the capabilities of artificial intelligence, this study utilizes neural networks to depict the correlations between US and International Public Finances and predict the changes in international public finances based on the changes in US public finances. With the neural network model achieving a commendable Mean Squared Error (MSE) value of 2.79, it is able to affirm a discernible correlation and also plot the effect of US market volatility on international markets. To further test the accuracy and significance of the model, an economic analysis was conducted that aimed to correlate the changes seen by the results of the model with historical stock market changes. This model demonstrates significant potential for investors to predict changes in international public finances based on signals from US markets, marking a significant stride in comprehending the intricacies of global public finances and the role of artificial intelligence in decoding its multifaceted patterns for practical forecasting.
\end{abstract}

\begin{IEEEkeywords}
computational finance, artificial intelligence, international markets, economics, spatiotemporal data analysis 
\end{IEEEkeywords}

\section{Introduction}
From the United States Treasury to financial administrations in capital cities across the world, the machinery of governance is fueled by immense flows of public sector debt and assets \cite{PFTheory}. Public finance encompasses these intricate governmental financial systems that enable vital public services, investments in national projects, and broader economic functions fundamental for societal stability \cite{PublicFinance}\cite{PFTheory2}. 

At its core, public finance is the financial activities and decisions made by the government to run the country and provide services to its citizens \cite{DynamicPF}. Therefore, it represents a balancing act: governments must judiciously administer assets generated from tax revenues and manage liabilities in the form of bond issuances or deficits that fund public welfare projects \cite{PFEcon}.
Historically, public finances remained relatively siloed within the confines of each nation state \cite{SocioApproach}. Balance sheets tallying streams of tax revenues and towering liabilities from bond issuances mostly fluctuated from internal dynamics. However, in today’s intensely integrated global economy, even minor changes in one country’s public debt markets can trigger a butterfly effect that links to fiscal conditions across international bounds \cite{Europe}. As the acceleration of globalization unravels traditional borders, understanding the emergence of these correlations in public finances becomes critical as it can help investors predict changes to the international markets based on changes in the US markets \cite{PFinGrowth}\cite{PFinPChoice}.

The capabilities of artificial intelligence methods have grown enormously in recent years, now powering innovative analyses that are rippling across the finance industry \cite{MLFinance}. Mathematical and statistical AI approaches can unpack complex patterns in vast financial data sets that evade traditional techniques \cite{New_perspective}. In particular, neural networks have become a highly useful AI method for gleaning insights from complex, multidimensional financial data. Neural networks can detect intricate nonlinear relationships and temporal dynamics within and between equities, bonds, currencies, and other financial assets. Their flexible modeling of intricate real-world patterns in finance data serves neural networks as an invaluable addition to the AI toolkit revolutionizing 21st century financial analysis.

Therefore, in this research I utilize artificial intelligence to conduct an analysis of stock change between US and International Public Finances. By employing a neural network, I am able to depict the correlations between US and International Public Finances and predict the changes in international markets based on the changes in US public finances.
The model achieved a Mean Squared Error (MSE) value of 2.79, proving a correlation between the two. Furthermore, through an economic analysis I am able to correlate the changes seen by the results of the model with historical stock market changes, overall demonstrating the significant potential of this model for investors to predict changes in international public finances based on signals from US markets.

\section{Materials and Methodology}
\subsection{Materials}
In this study, I utilized credit rating history of US Public Finances and International Public Finances from S\&P Global, since November 1, 2010. While the datasets consisted several metrics, the data points utilized in the analysis were the initial credit ratings, change in credit ratings, and the date of change for the all entities examined. 
\subsection{Data Preprocessing}
To facilitate meaningful comparisons and analyses, a meticulous literature review was conducted for feature engineering. Recognizing the temporal dimension as a significant factor, the primary approach involved leveraging the credit rating change dates as the chronological anchor for the analysis. This strategic use of time allowed for seamless alignment of data points from the US and International markets.

The datasets were then concatenated based on the timeline, ensuring the creation of a cohesive and comprehensive main dataset for subsequent analysis. This temporal alignment formed the foundation for understanding and comparing the dynamics of the credit ratings.

Moreover, during data preprocessing, missing values and outliers were addressed through careful imputation and normalization techniques. This ensured the integrity of the dataset and the robustness of subsequent analyses. Exploratory data analysis was also conducted to gain insights into the distribution and characteristics of the selected metrics, providing a crucial context for the subsequent application of artificial intelligence techniques.

By establishing a well-organized and temporally aligned dataset through rigorous data preprocessing, the subsequent artificial intelligence-based analysis was poised to effectively capture and interpret the intricate relationship between changes in US financial data and their impact on international finances. 

\subsection{Artificial Intelligence}
In the realm of artificial intelligence, the pivotal aim was to construct an algorithm that could intricately predict international public finance ratings based on US public finance ratings. The foundational element of this AI-based analysis was the integration of a Long Short-Term Memory (LSTM) based input layer \cite{UnderstandingLSTM}. This specialized layer was meticulously designed to assimilate and comprehend the nuanced intricacies embedded within the financial datasets. Notably, the LSTM played a crucial role in unearthing temporal correlations, allowing the model to discern evolving patterns over time \cite{UnderstandingNN}.

The subsequent layer in the neural network architecture comprised a singular node within the output layer, responsible for predicting changes in international finances based on the fluctuations observed in US finances. The LSTM's significance extended to facilitating an understanding of correlations between input and output at specific temporal junctures \cite{NNApplications}. This temporal awareness proved essential in capturing causal relationships across multiple time steps, enriching the model's ability to discern the intricate dynamics of financial changes \cite{LSTM}.

To further augment the feature extraction process and enable intricate nonlinear mapping, a dense layer was strategically introduced \cite{Gers2000}. This layer, adept at extracting key features from the data, engaged in complex nonlinear mappings between these extracted features and the target output \cite{Graves2005}. The relationship between the LSTM and dense layers culminated in the creation of a sophisticated neural network architecture. This architecture exhibited a remarkable proficiency in comprehending the intricate relationships within time series data.

Through this strategic combination, the neural network not only effectively mapped changes in US financial data but also demonstrated a remarkable capacity to predict corresponding shifts in international finances. The model's prowess in capturing spatiotemporal correlations validated the efficacy of the chosen AI methodology, showcasing its potential to provide nuanced insights into the complex dynamics underlying the relationship between US and international public finances.

\section{Results}
\begin{figure}[!htb]
	\centering
	\includegraphics[width=\columnwidth]{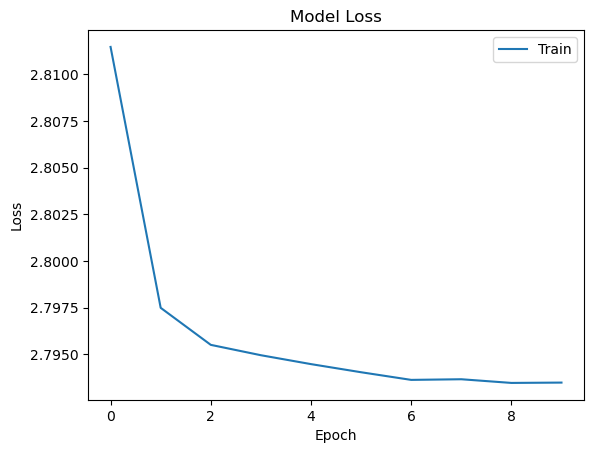}
	\caption{Loss Function as Neural Network Learns}
	\label{fig1}
\end{figure}

To measure the accuracy of the neural network one key metric was utilized: the Mean Squared Error(MSE). The MSE that the neural network was able to achieve was 2.79, indicating that the neural network was able to predict the change that was happening in the international stocks based on the US stocks with high accuracy. Furthermore, to see how well the neural network was learning the data it was important to look into the way the neural network was learning the data throughout the epochs. The neural network's ability to learn the data can be seen in Figure \ref{fig1}. 

The data graph was created through detailed statistical analysis which clearly showed major financial market changes over recent years. By examining the graphs and charts, I saw subtle indications in metrics and trends preceding these shifts. For example, small volatility upticks often foreshadowed larger spikes and turbulence. Similarly, gradual trading volumes and frequency pattern changes exposed developing momentum before rapid price rises or falls. Though major shifts seem sudden, analysis shows small tremors often presage them, representing building pressure within the system. Identifying those early signs allows one to predict and prepare for coming movements. So by creating data visualizations and analyses, I illuminated complex indicators preceding and prompting the most significant changes.

\FloatBarrier
\begin{figure*}[!htb]
	\centering
	\includegraphics[width=1.5\columnwidth]{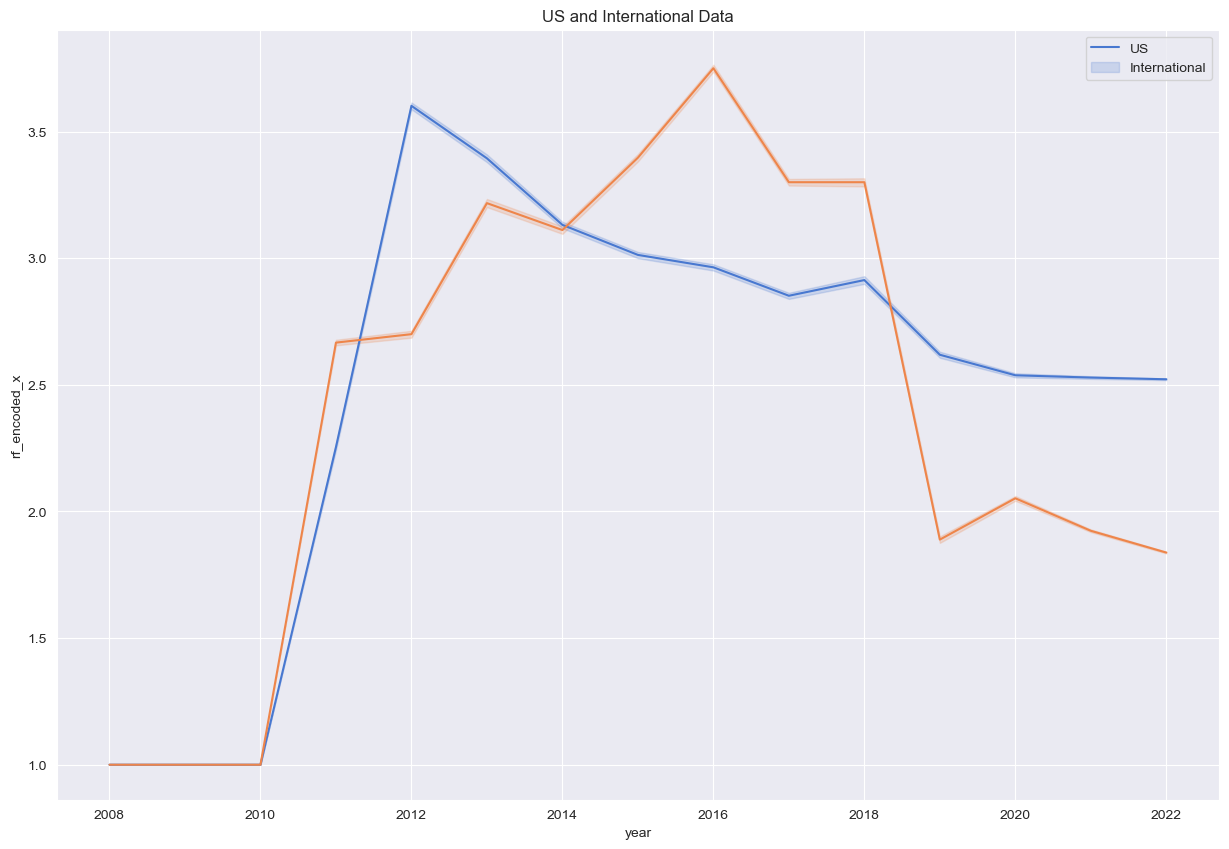}
	\caption{US vs. International Markets}
	\label{fig2}
\end{figure*}
\FloatBarrier

\section{Economic Analysis}
Based on Figure \ref{fig2}, we are able to see the robust result of the model in terms of analyzing the correlation between the changes in public finances among US and International entities, and specifically how these international entities react to changes within the US. However, to verify this trend, we can compare the change in the US Public Finances to major stock events that occurred in the past 10 years, to check if the model indeed matches and see the correlating change the events have on international markets in terms of how much stocks changed and how long it took for the international markets to reflect those changes.

Starting with the first major change we see in Figure \ref{fig2}, which takes place in 2011, can be credited to the stock market crash on August 8, 2011. In August 2011, the stock market plunged largely due to a cascading crisis of confidence among investors triggered by mounting debt fears in Europe and a stall in US economic growth recovery \cite{Jayech2016}. Specifically, the downgrade of the US credit rating, ongoing failure by US and European leaders to contain the Eurozone crisis, and skittish reactions from regulators created a panic. Investors frantically sold off stocks across global exchanges, withdrawing over \$15 billion in assets within days and shifting capital to perceived safer investments like treasury bonds. The benchmark S\&P 500 fell 10.7\% between July 22 and August 8. Overall, trillions in equity value evaporated within weeks due to this macroeconomic uncertainty combined with political paralysis, uninspiring economic fundamentals, and unreliable growth projections from corporations. The plunging markets reflected overwhelmed central banks, credit contraction, recession fears, and loss of investor faith in the system to properly value assets and risks \cite{Politics}. As seen in Figure \ref{fig2}, this crash is represented by the change in the US Public Finances, with the change in International Public Finances following at the beginning of 2012.  

In 2013, we see a similar decline in the US public finances, which is likely a result of the 2013 federal government shutdown. The 2013 federal government shutdown was a political budget crisis that resulted from the failure of Congress to pass regular appropriations bills required to fund federal government operations \cite{Baker2017}. The shutdown lasted 16 days in October 2013 due to an impasse between Republican and Democrat lawmakers on government spending. During the shutdown period, many routine governmental services halted, federal sites closed, hundreds of thousands of nonessential federal employees were furloughed, and spending on contracts to businesses deferred. This disruption significantly lowered income, corporate, payroll and other tax revenues flowing into public finances for funding governance. Additionally, the shutdown provoked fiscal uncertainty that introduced volatility in financial markets and dampened projections for economic growth- also limiting prediction reliability for cyclically-sensitive tax obligations. Overall the shutdown was estimated to curb GDP by \$24 billion and substantially cut near-term tax inflows into public finances. Thus the ripple effects from the 2013 political crisis likely translated into a pronounced dip and deceleration in overall US public finances accumulation that year- both directly from the temporary suspension of taxation streams, and indirectly via market and economic impacts eroding revenue performance tied to the shutdown. This dip likely could have also been a result of the 2013 NASDAQ Flash Freeze. On August 22, 2013, an unprecedented three-hour outage froze trading on the NASDAQ stock exchange, suspending all transactions and paralyzing the second-largest stock market in the US \cite{kern2018high}. With over \$25 billion worth of transactions affected and market confidence rattled, the technical glitch introduced enough fiscal and economic uncertainty that year to likely translate into a pronounced dip in income, capital gains, and investor sentiment-driven tax revenues for U.S. public finances. As seen by the international trend line, this crash was reflected at the beginning of 2014.

In Figure \ref{fig2}, we also see another dip in the year of 2015-16, which was likely a result of the 2015-16 stock market selloff, a period of decline in the Dow Jones Industrial Average, S\&P 500 Index, and NASDAQ Composite that lasted from August 18, 2015 until February 11, 2016 \cite{so2022assessing}. Triggered by fears of a slowdown in China and plunging oil prices, US stocks lost over 10\% of value in just 6 days in August 2015, indicative of a market correction veering into bear market territory. While US economic fundamentals remained sound, the selloff eroded investor confidence and introduced volatility that persisted for months. With over \$3 trillion in global equity value shed, this historic selloff likely stunted predicted profit levels, investment plans, and employment expansions upon which cyclical tax revenue projections for public finances rely. The bearish sentiment and selloff’s ripple effects likely slowed multiple taxation income streams from Wall Street speculation cuts to Main Street economic impacts– translating into a pronounced dip or disruption in accumulating public funds during this 2015-2016 period. This crash was reflected almost immediately by the international markets.

The dip seen in 2018, is indicative of the 2018 cryptocurrency crash, which was a massive selloff in the cryptocurrency markets that accelerated in January 2018, as major digital assets like Bitcoin and Ethereum plunged over 60\% from record highs set in late 2017 \cite{manahov2023great}. Catalyzed by new regulations, security breaches, and Facebook banning crypto ads, the crash wiped out over \$300 billion in total cryptocurrency market capitalization by 2018’s end. With cryptocurrency investment and trading subject to capital gains tax obligations, this extreme market correction likely depressed fiscal year tax payments from digital currency investors declaring losses or at least hugely reduced gains versus projections. Additionally, diminished investor appetite filters into sales, payroll, and corporate taxes if industry contraction or failures transpire. While still an emerging asset class, crypto’s high growth suggested the 2018 crash introduced enough economic and fiscal uncertainty amid plunge warnings that a slight dip in certain US public finance revenue streams tied to cryptocurrency could have emerged. This crash was also reflected immediately by the international markets.

Finally, the last dip seen in 2020 represents the fall as a result of the Coronavirus pandemic. The 2020 Coronavirus pandemic was a global viral disease outbreak caused by the highly contagious COVID-19 virus first identified in Wuhan, China in late 2019. As the World Health Organization declared a Public Health Emergency in January and pervasive spread prompted a worldwide economic shutdown by March 2020, US public finances were profoundly disrupted \cite{mazur2021covid}. Mass business closures, supply chain failures, and over 20 million jobs lost within months slowed economic activity to Depression-era contraction levels and eviscerated tax revenue inputs from income to retail sales to hospitality and transportation. With interest rates slashed to zero, Congress approved over \$3 trillion in emergency pandemic stimulus and bailouts by mid-2020 - massively bloating the Federal budget deficit towards unprecedented peacetime levels over 15\% of GDP. This economic “sudden stop” coupled with tremendous relief spending created largest shock to the accrual and outflows of US public finances since World War II, with multi-trillion dollar budget gaps persisting years later. This dip was also reflected immediately by international markets.

\section{Implications}
The ability to model international public finance dynamics based on signals from US markets demonstrated by this neural network carries noteworthy practical implications. Investors and policymakers alike stand to benefit from the predictive insights and risk analysis enabled by the model’s skill in mapping cross-border correlations.

Specifically, the model can assist investors in public finance instruments like government bonds in better diversifying systemic risks in these concentrated, overarching debt markets that shape broader asset valuations. If US fiscal or monetary policies portend turbulence that will propagate internationally, this intelligence allows tactical portfolio adjustments for international managers. The model may also have predictive power for relative currency and inflation effects. Policymakers can likewise employ these projections to make nuanced stimulus or austerity decisions factoring global impacts.

More broadly, the realized interconnectivity surfaces policy vulnerabilities like the potential for financial contagions during crises. This understanding may promote productive cooperative forums between the US and particular foreign nations to coordinate interventions, regulations, or structural reforms that mitigate harmful spillovers. However, the throughline between once discrete public finance systems could also compound tensions between national and international economic interests.

This neural network’s technical methods constitute an early stride in a burgeoning subfield of computational intelligence for fiscal and financial analyses. The promising proof of concept invites streamlining for real-time deployment and pairing with causal inference techniques to enrich interpretability.

\section{Conclusion}
This study demonstrated the efficacy of using artificial intelligence, specifically LSTM neural networks, to model intricate correlations between shifts in US public finances and subsequent impacts on international markets. The model achieved robust predictive accuracy with a low MSE of 2.79, capturing discernible relationships in the temporal data. 

The analysis of historical periods of volatility further validate the model’s learned linkage between precipitous changes in US public debt instruments or tax revenues and ensuing global financial impacts. Periods like the 2011 Stock Market Crash, 2013 US Government Shutdown, 2015-16 Stock Selloff, 2018 Crypto Crash, and 2020 COVID Crisis exhibited pronounced co-movement between US fiscal disruptions and international finance, affirming the model’s technical signal amidst the noise.

The practical utility of this predictive capacity for both policymakers and investors is noteworthy. International debt and currency markets remain concentrated systemic risk points for global growth. This neural network’s ability to foresee how US fiscal and monetary actions may propagate abroad grants a key instrument for scenario analysis, risk evaluation for international assets, and policy coordination.

While this study marks an initial foray applying machine learning to decode integrated public finance systems, the promising proof of concept warrants significant future work. Scalability to many other national economies, streamlining for real-time automated outputs, deepening interpretability, and pairing predictive fiscal AI with policy decision support systems offer rich potential. As globalization compounds interconnectedness, computational intelligence in tracing financial reverberations across borders grows ever more indispensable.

\section{Future Works}
While this research pioneered an AI methodology tracing fiscal interconnectedness between the US and international markets, it remains an initial foray warranting extensive future efforts as follow-ons. 

One natural progression involves expanding model coverage from the singular US economy to multiple other mature economies and fiscal union entities. Tailoring additional country-specific LSTM input layers would paint a more global portrait of public finance co-movements and enhance early warning systems for cross-border financial contagion threats.  

Another critical next step entails streamlining the model for real-time automated outputs leveraged in live fiscal monitoring systems. This could enable policymakers to gauge international market reactions to major domestic spending bills or tax policy shifts before enactment. High-frequency trading firms may also exploit these efficient predictions in their algorithmic systems.

Further refinement of the neural network with supplementary techniques promises to improve result interpretation from correlation to precise causation tracing. Incorporating explainability approaches like SHAP values into model post-processing can isolate the magnitude of importance of specific US policy actions on international outcomes. These enriched insights strengthen multi-lateral coordination efforts.

Lastly, integrating state-of-the-art predictive fiscal AI directly into policy decision support architectures could unlock tremendous upside. Beyond passive market estimations, optimizing stimulus and taxation levels to navigate global growth shockwaves through automated intelligence represents an intriguing continuation. The possibilities are as boundless as innovation at the intersection of machine learning and public financial administration in the 21st century global economy.

\section{Acknowledgment}
I would like to thank my professor Dr. Stephen Owen at the University of North Texas for his kind and thoughtful guidance and mentorship throughout this research study.

\bibliographystyle{IEEEtran}

\begin{thebibliography}{99}

\bibitem{PublicFinance}
Rosen, H.S., \textit{Public Finance}. The Encyclopedia of Public Choice, Springer US, Boston, MA, 2010. 


\bibitem{Europe}
Romero-Ávila, D., Strauch, R., Public finances and long-term growth in Europe: Evidence from a panel data analysis. \textit{European Journal of Political Economy} \textbf{24}(1), 172-191 (2008). 

\bibitem{DynamicPF}  
Kocherlakota, N.R., \textit{The New Dynamic Public Finance}. Princeton University Press (2010). 

\bibitem{SocioApproach}
Goldscheid, R., A Sociological Approach to Problems of Public Finance. In: \textit{Classics in the Theory of Public Finance}. Palgrave Macmillan UK, London (1958). 

\bibitem{PFinGrowth}
Barro, R.J., Sala-I-Martin, X., Public Finance in Models of Economic Growth. \textit{The Review of Economic Studies} \textbf{59}(4), 645 (1992). 

\bibitem{PFinPChoice}  
Buchanan J.M., Public Finance and Public Choice. \textit{National Tax Journal} \textbf{28}(4), 383-394 (1975). 

\bibitem{PFEcon}
Persson, T., Tabellini G., Political Economics and Public Finance. In: \textit{Handbook of Public Economics} (2002). 

\bibitem{PFTheory}
Wagner A., Three Extracts on Public Finance. In: \textit{Classics in the Theory of Public Finance}. Palgrave Macmillan UK, London (1958). 

\bibitem{PFTheory2}
Oates W.E., The Theory of Public Finance in a Federal System. \textit{The Canadian Journal of Economics} \textbf{1}(1), 37 (1968). 

\bibitem{Politics}
Persson, T., Roland G., Tabellini G., Comparative Politics and Public Finance. \textit{Journal of Political Economy} \textbf{108}(6), 1121-1161 (2000).

\bibitem{New_perspective}  
Stiglitz, J.E., New perspectives on public finance: recent achievements and future challenges. \textit{Journal of Public Economics} \textbf{86}(3), 341-360 (2002). 

\bibitem{UnderstandingNN}
Warner, B., Misra, M., Understanding Neural Networks as Statistical Tools. \textit{The American Statistician} \textbf{50}(4), 284-293 (1996). 

\bibitem{NNApplications}
Bishop, C.M., Neural networks and their applications. \textit{Review of Scientific Instruments} \textbf{65}(6), 1803-1832 (1994). 

\bibitem{MLFinance} 
Ghoddusi, H. Creamer G.G., Rafizadeh N., Machine learning in energy economics and finance: A review. \textit{Energy Economics} \textbf{81}, 709-727 (2019). 

\bibitem{UnderstandingLSTM}
Staudemeyer, R.C., Morris, E.R., Understanding LSTM – a tutorial into Long Short-Term Memory Recurrent Neural Networks (2019).

\bibitem{LSTM} 
Sherstinsky A.,  Fundamentals of Recurrent Neural Network (RNN) and Long Short-Term Memory (LSTM) network. \textit{Physica D: Nonlinear Phenomena} \textbf{404}, 132306 (2020). 

\bibitem{Gers2000}
Gers F.A., Schmidhuber J., Cummins F., Learning to Forget: Continual Prediction with LSTM. \textit{Neural Computation} \textbf{12}(10), 2451-2471 (2000). 

\bibitem{Graves2005}
Graves, A., Schmidhuber, J., Framewise phoneme classification with bidirectional LSTM and other neural network architectures. \textit{Neural Networks} \textbf{18}(5-6), 602-610 (2005).

\bibitem{Jayech2016} 
Jayech S., The contagion channels of July–August-2011 stock market crash: A DAG-copula based approach. \textit{European Journal of Operational Research} \textbf{249}(2), 631-646 (2016). 

\bibitem{Baker2017}
Baker, S.R., Yannelis, C., Income changes and consumption: Evidence from the 2013 federal government shutdown. \textit{Review of Economic Dynamics} \textbf{23}, 99-124 (2017). 

\bibitem{kern2018high}
Kern, S., Loiacono, G., High Frequency Trading and Circuit Breakers. In: \textit{Global Algorithmic Capital Markets: High Frequency Trading, Dark Pools, and Regulatory Challenges}. Oxford University Press (2018).

\bibitem{so2022assessing} 
So, M.K.P., Mak, A.S.W., Chu, A.M.Y., Assessing systemic risk in financial markets using dynamic topic networks. \textit{Scientific Reports} \textbf{12}(1), 2668 (2022).  

\bibitem{manahov2023great}
Manahov, V., The great crypto crash in September 2018: why did the cryptocurrency market collapse? \textit{Annals of Operations Research}, 1-38 (2023).  

\bibitem{mazur2021covid}
Mazur, M., Dang, M., Vega M., COVID-19 and the march 2020 stock market crash. Evidence from S\&P1500. \textit{Finance research letters} \textbf{38}, 101690 (2021).  


\end{thebibliography}

\end{document}